\documentclass[twocolumn]{article}

\usepackage{amsmath,amsfonts,epsfig,graphics,amssymb}
\usepackage{cite}
\usepackage{hyperref}
\usepackage{authblk}
\usepackage[margin=1.8cm, top = 2.2cm, bottom=3cm]{geometry}
\hypersetup{colorlinks=true,citecolor=blue}

\title{\bf First measeurements in search for keV-sterile neutrino in tritium beta-decay by Troitsk nu-mass experiment}
\author[1]{J.\,N.~Abdurashitov}
\author[1]{A.\,I.~Belesev}
\author[1]{V.\,G.~Chernov}
\author[1]{E.\,V.~Geraskin}
\author[1]{A.\,A.~Golubev}
\author[1,2]{P.\,V.~Grigorieva}
\author[2]{G.\,A.~Koroteev}
\author[1]{N.\,A.~Likhovid}
\author[1,2]{ A.\,A.~Nozik}
\author[1]{V.\,S.~Pantuev}
\author[1]{V\,.I.~Parfenov}
\author[1]{A.\,K.~Skasyrskaya}
\author[1,3]{I.\,I.~Tkachev}
\author[1]{S.\,V.~Zadorozhny}

\affil[1]{\small \it Institute for Nuclear Research RAS, 117312 Moscow, Russia}
\affil[2]{\it Moscow Institute of Physics and Technology, Dolgoprudny, Moscow region 141700, Russia}
\affil[3]{\it Physics Department, Novosibirsk State University, Novosibirsk 630090, Russia}

\renewcommand\refname{}
\date{}
\begin{document}

\twocolumn[
\begin{@twocolumnfalse}
\maketitle
\begin{abstract}
{We present the first results of precision measurements of tritium $\beta$-decay spectrum in the electron energy range  16--18.6 keV by the Troitsk nu-mass experiment. The goal is to find distortions which may be caused by the existence of a heavy sterile neutrinos. A signature would correspond to  a kink in the spectrum  with characteristic  shape and end point shifted by the value of a heavy neutrino mass. We set a new upper limits to the neutrino mixing matrix element $U^2_{e4}$ which improve  existing limits  by a factor from 2 to 5   in the mass range 0.1--2 keV.}
\vspace{\baselineskip}
\end{abstract}
\end{@twocolumnfalse}
]


\vspace{\baselineskip}

\paragraph{1. Introduction.}
Neutrino studies nowadays belong to the most active and promising research domain in particle physics.  Neutrinos are massive and this property cannot be accommodated in the Standard model. Therefore, studies of the neutrino mass matrix give us unique opportunity to probe a new physics  in the laboratory directly. The simplest mechanism to provide neutrino with the mass assumes the existence of a new particles - right handed (or "sterile") neutrinos. Their number, masses and mixing angles with left handed ("active") neutrinos, apart from the  existing observational restrictions, are free parameters of the theory,  for the review see \cite{Adhikari:2016bei}.

It is important that sterile neutrino in the keV mass range is one of the best motivated \cite{Adhikari:2016bei} dark matter particle candidates. Therefore, sterile neutrino searches in this mass range are of particular interest. 

Sterile neutrinos do not have Standard Model interactions (hence the name "sterile"). However, since sterile neutrinos mix with active neutrinos, they are emitted with small probability in every process where active neutrinos can be born. In particular,  in the presence of sterile neutrino the spectrum of tritium $\beta$-decay is modified in the following way 
\begin{equation}
	{S(E) =  U^2_{e4}S(E,m_4^2)+(1-U^2_{e4})S(E,m_{1}^2),}
\label{eq:funct}
\end{equation}
where $S(E,m)$ is standard $\beta$-spectrum  with single neutrino if it has mass $m$; $U_{e4}$ is relevant element of the 
neutrino mixing matrix; mass states $m_4$ and $m_1$ practically coincide with sterile and electron neutrino correspondingly,  if mixing is small. 
Sterile neutrinos can be searched  for by looking for such small $\beta$-spectrum distortions.

Troitsk nu-mass experiment, which had been designed for the active neutrino mass measurements, and where the strongest direct limit on it  \cite{Aseev:2011dq,Kraus:2004zw} has been obtained so far,  is well suited for the above task~\cite{Abdurashitov:2015jha}. Moreover, currently, it is the only installation in operation which is capable for the sterile neutrino searches in the keV mass range. We have started corresponding research program recently.    
This Letter reports  on the first physical results and restrictions obtained  in searches for sterile neutrinos in the mass range up to a few keV. 

\paragraph{2. Troitsk nu-mass.}

Detailed description of Troitsk nu-mass experiment, hardware, procedures and  important  systematic errors inherent to the spectrum measurements on this apparatus, can be found in Ref.~\cite{Abdurashitov:2015jha}. 

In short, we are measuring integral spectrum of tritium $\beta$-decay. 
Our equipment consists of two main components which are the Windowless Gaseous Tritium Source (WGTS) and the Electrostatic Spectrometer with Magnetic Adiabatic Collimation (MAC-E filter). Gaseous source allows to avoid solid state effects which distort spectra if  tritium is frozen or implanted.  The radioactive gas freely circulates in a source and there are no effects associated with a "wall" or "window". Electrons originating in the tritium decay are transported adiabatically by the system of superconducting solenoids to the spectrometer entrance with acceptance of 2.1\% which is defined by magnetic filed values in the WGTS and the spectrometer magnet. Within our energy range acceptance variation is negligible.  
MAC-E filter works as the electrostatic rejector of all electrons with energy less than a given spectrometer potential, regardless of initial direction of electron momentum within acceptance. All electrons with higher energy are counted by the detector. Measurements with different values of retarding potential  give integral  energy spectrum of electrons. 

We stress that in our experiment it is sufficient to just count electrons, however, that should be done efficiently. 
We count electrons by a single channel Si(Li) detector with 30 nm palladium window and Si 100 nm dead zone. Signals from the detector are shaped by the amplifier-shaper with 2 microseconds integration time and then  digitized  by a system with constant (approximately $6.5$ microseconds) dead time. Such a long integration time is required in order  to minimize the back scattering effect of electrons from the detector, which reaches 10-15\% probability. Then these scattered electrons are reflected with some delay by the electrostatic and magnetic mirrors back to the detector. Relevant details of this feature of the MAC-E filter are described in Ref.~\cite{Grigorieva:2015wkq}, while  systematic errors related to the detection efficiency are described, in Section 4 below.

\paragraph{3. Measurements.}

Since we are searching for the spectrum distortions caused by possible emission of sterile neutrinos, all other sources of distortions should be minimized.  Therefore, the spectrum measurement procedure has been fixed to the following scheme. 

The spectrometer electrostatic potential was scanned from 18700~V (which exceeds the end point of the spectrum located at 18570~eV) down to 16000~V,\footnote{This threshold was defined by the maximum counting rate in our current configuration. We will do measurements in a wider range and at lower potentials in the future measurements.} with steps of 50~V. 
At each point the high voltage was controlled with the precision of about $\pm$0.2~V and measurements were taken during 30 seconds. To avoid possible distortion of the spectrum caused by the temporal instabilities  we scan in series of high voltage going down and up. In addition, to control the stability of the radioactive tritium source, point at 16000~V was selected as the monitor point and was measured after every 6-8 steps. 

As a working gas we have used DT molecules with small concentration of HT and T$_2$. The total column density of the tritium source corresponded to approximately 10\% probability of electron scattering in the gas and was periodically measured employing the electron gun located at the rear side of the WGTS. To stabilize the amount of working gas in the WGTS pipe, its temperature was adjusted to 26-28~K by the combination of cold helium gas from the cryogenic system and additional heater. During each set of measurements the pipe temperature was stable with precision better than 0.1~K. The maximum  rate  of electron counting on the detector was reaching 15 kHz at 16 kV  spectrometer potential. One gas fill lasted for 5-6 days. At the end of a fill the intensity of the tritium source was dropping by a few tens of a percent.

At the end of a fill, when the working gas was pumped out, we performed control measurements of the empty source with four spectrometer cycles up and down. That was used to subtract physical background.

In addition, precise measurements of the spectrometer transmission function were done using electron gun operating at energies between 16~keV and 19~keV. Transmission function deviates from the step function characterizing an ideal spectrometer due to effect described in~\cite{Grigorieva:2015wkq}, and stronger effect arises  at  potentials  lower than 14~kV (in the range which we do not employ in the present measurements) due to break up of adiabaticity.   We used e-gun calibration results to describe and calibrate spectrometer response function.

\paragraph{4. Corrections and systematics.}

The expected list of systematic uncertainties in our experiment is presented in Ref.~\cite{Abdurashitov:2015jha}. 
At the current level of accumulated statistics the most important sources of systematic uncertainties are: events under detection threshold and electronics dead time. As we have mentioned already, our detector is working in the electron counting regime, but the energy dependence of counting efficiency, to which both effects listed above contribute, should be understood and corrected. 

\begin{figure}
	\begin{center}
		\includegraphics[width=.95\linewidth]{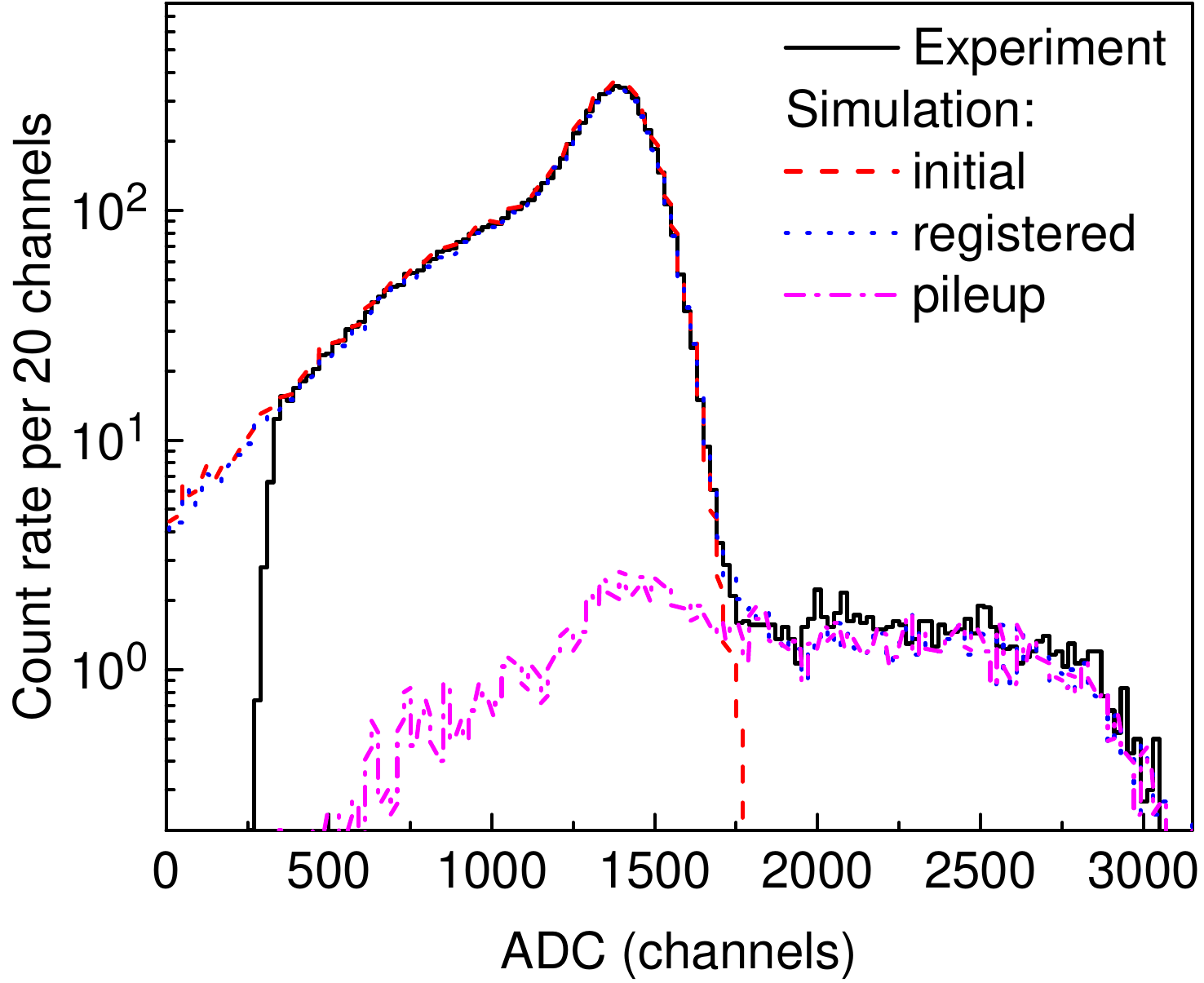}
	\end{center}
	\caption{Amplitude spectrum of the detector at the spectrometer potential fixed to 16.0~kV. Solid line corresponds to the measured spectrum. Red dashed line represents the initial distribution used in simulation.  Dotted  line corresponds to the simulated spectrum. Dot-dashed line shows extracted contribution of pileup events.}
	\label{fig:pileup}
\end{figure}

Amplitude spectrum of our detector  obtained in the real  measurements  of the tritium $\beta$-decay is shown in Fig.~\ref{fig:pileup} by the solid line. Spectrometer potential was fixed to 16.0~kV. This spectrum contains two artifacts which should be corrected. First is the lack of events below ADC channel $\approx 250$ caused by detection threshold. Second is the tail above $\approx 1750$ due to event pileup caused by  electronics dead time. Below we describe procedure which we have used to correct for both of these effects inherent to any detector. 

{\bf  4.1. Events under threshold.}
We register electrons which produce signal  with amplitude above some threshold, the latter was set by electronics to cut down the noise. The loss of undetected electrons should be corrected for.  We do this by extrapolating the amplitude spectrum into the region below threshold (and down to zero in counting rate).

Measured amplitude spectrum  follows exponential distribution at low amplitudes, see~Fig.~\ref{fig:pileup}.   Simulation of the back scattering processes from the detector with reflection of the scattered particles back to the detector by  electrostatic and magnetic mirrors~\cite{Grigorieva:2015wkq}, which dominate in this region of amplitudes, qualitatively shows similar exponential shape. The measured slope of this exponent is almost the same for different spectrometer potential or the electron energy. We extrapolate the ADC spectra using exponential fits in the narrow region  400 -- 600 ADC channels, and estimate the number of events under the threshold as an integral of this exponential function. The correction could not be calculated in this way  for higher retarding potentials (above 17.5~kV), where the number of events is insufficient to do a reliable fit. Therefore, correction itself was extrapolated into this region using its shape at lower potentials, which is also exponential.
The final correction for the threshold effect is 2-4~\%. We assume that the systematic error here comes from the uncertainty of the fitting procedure.

{\bf  4.2. Dead time.}
Currently the maximum count rate at the detector is 14-16 kHz and is limited by the $\sim6.5~\mu s$ electronics dead time. If two events are close enough, namely in the range of about 2.5~$\mu s$, we get ``pileup event" which is indistinguishable from a single event, but creates larger registered amplitude of the signal. 
In order to disentangle  pileup and to find real counts, $N_{real}$, we made a Monte-Carlo simulation using the  following calibration data:
\begin{itemize}
	\item  dependence of the electronics dead time  on the signal amplitude, measured by the two-signal pulser with adjustable distance between signals;
	\item pileup probability and amplitude for different delays using the same pulser;
\end{itemize}

The comparison of simulated and real detector responses is presented in Fig.~\ref{fig:pileup} and shows very good agreement between them. 	Simulation started with the initial distribution which was close to the  amplitude spectrum of  real events in the experiment,  but with slightly larger intensity of count rate, and without  pileup, see red curve in Fig.~\ref{fig:pileup}. Integral under this  curve gives desired $N_{real}$,  while integral under experimental curve we denote  $N_{det}$ in what follows.

\begin{figure}
	\begin{center}
		\includegraphics[width=.95\linewidth]{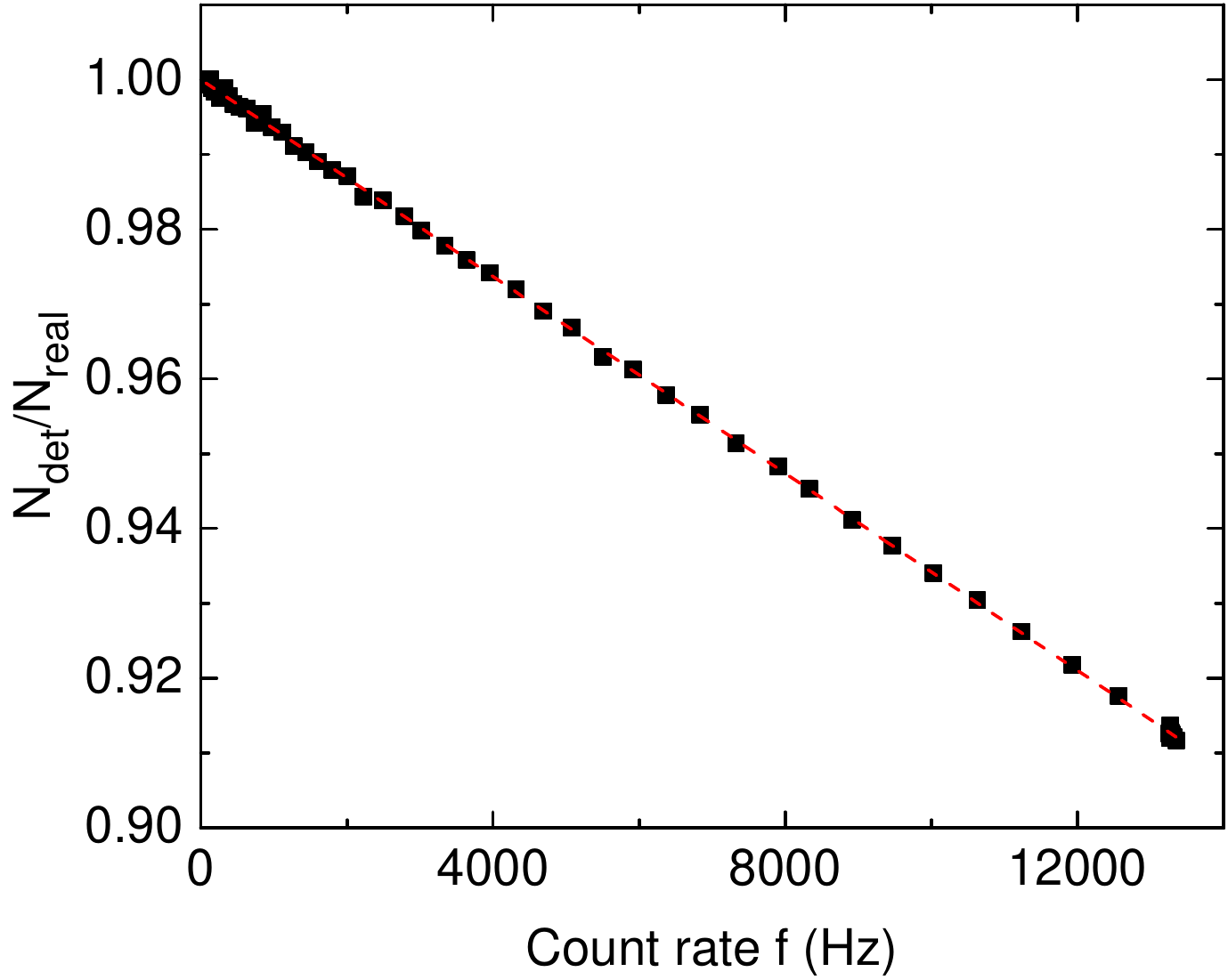}
	\end{center}
	\caption{Dead time correction factor as a function of $f$.}
	\label{fig:efficiency}
\end{figure}

Simulation was repeated for each spectrometer potential using real amplitude spectra as a template.  Resulting dependence of  $N_{det}/N_{real}$ on the count rate is shown  in the Fig.~\ref{fig:efficiency}, and represents dead time correction factor.

This dependence can be described by the formula:
\begin{equation}
	N_{real} = \frac{N_{det}}{1-f\tau},
\label{eq:1}	
\end{equation}
where $f$ is {\it measured} count rate integrated over all channels and $\tau = 6.55 \pm 0.05~\mu s$ is the {\it effective} dead time, which was found by fitting Eq.~({\ref{eq:1}) to the  MC results shown in Fig.~\ref{fig:efficiency} by black boxes. This effective dead time should not be confused  with dead times found in the calibration procedure. The latter depend slightly on the amplitude of a signal, the former is proper average.

{\bf  4.3.  Other corrections.}

Besides electronics dead time and events under threshold we applied our usual corrections  which were already discussed in detail in Refs.~\cite{Aseev:2011dq,Abdurashitov:2015jha,Grigorieva:2015wkq}. Important corrections and corresponding effects are  listed below. 
\begin{itemize}
	\item  Trapping effect. We have included the amplitude  of the trapping effect as a free parameter in the analysis of the tritium beta-decay spectrum. 
	\item Spectrometer  transmission function. There is a small variation of the transmission in the energy range 16 -- 19~keV caused by the detector backscattering effect, and spectra were corrected assuming linear interpolation of transmission function between these points.
\end{itemize} 

There are also less important systematic effects, which were treated in the same way as in our previous work. 
Those are: correction for the final state spectrum of residual ion of the daughter molecule $D^3He^+$, instability of high voltage, and electron scattering in the source. 
 
\paragraph{5. Results.}

\begin{figure}[t]
	\begin{center}
		\includegraphics[width = .95\linewidth]{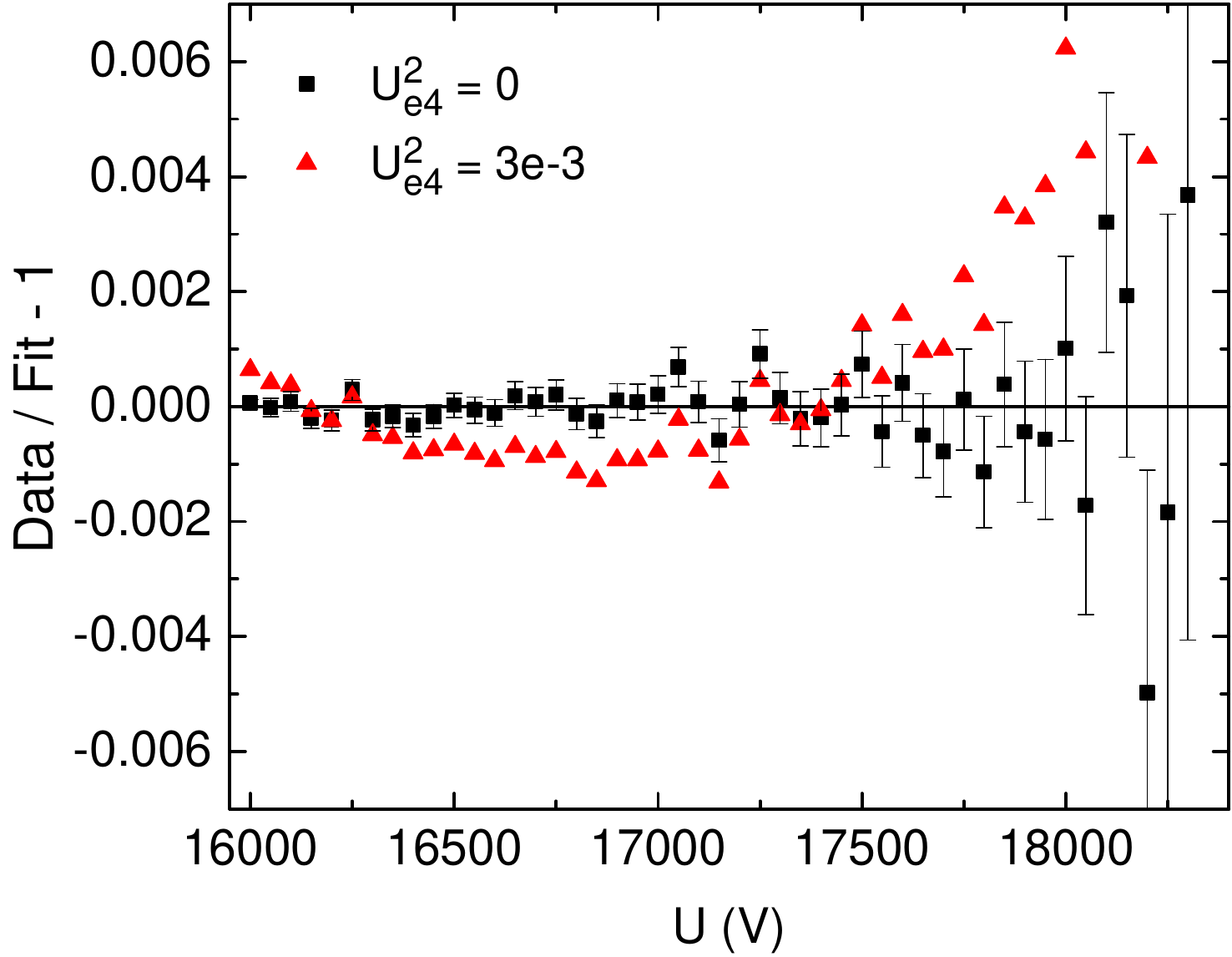}
	\end{center}
	\caption{Difference between measured $\beta$-spectrum and its best fit theoretical expectation. Black squares - model without sterile neutrino, red triangles - model with  1 keV sterile neutrino and mixing $U^2_{e4} = 3\cdot 10^{-3}$. To avoid clutter we display error bars for one model only.}
	\label{fig:spectrum}
\end{figure}

The final spectrum of tritium $\beta$-decay (after corrections) was fitted by the function Eq.~(\ref{eq:funct}). It is shown in Fig.~\ref{fig:spectrum} as residuals with theoretical expectations since achieved small errors and variations of spectral shape for different models cannot be seen on the spectrum itself. The model without sterile neutrinos (or zero mixing for any mass of sterile) perfectly fits  the data. Large mixing angles are not allowed. For example, the case of  1 keV sterile neutrino and mixing $U^2_{e4} = 3\cdot 10^{-3}$ is clearly excluded, see Fig.~~\ref{fig:spectrum}.

We have applied  sensitivity limit procedure (see ref. \cite{Lokhov:2014zna} for details) to derive  upper limits for allowed $U^2_{e4}$ . The result is shown in Fig.~\ref{fig:limits} as a function of sterile neutrino mass. 
In the same figure we also present the  existing published limits
~\cite{Kraus:2012he, Belesev:2012hx, Belesev2, Galeazzi:2001py, Hiddemann:1995ce}.

To conclude, we present our first results on the  search of a heavy sterile neutrinos in tritium $\beta$-decay   in the mass range of a few keV.   We have improved  the existing limits on their mixing with electron neutrinos by the factor of 2 to 5, depending upon $m_x$, in the mass range 0.1 -- 2 keV. 

\begin{figure}[t]
	\begin{center}
		\includegraphics[width=0.95\linewidth]{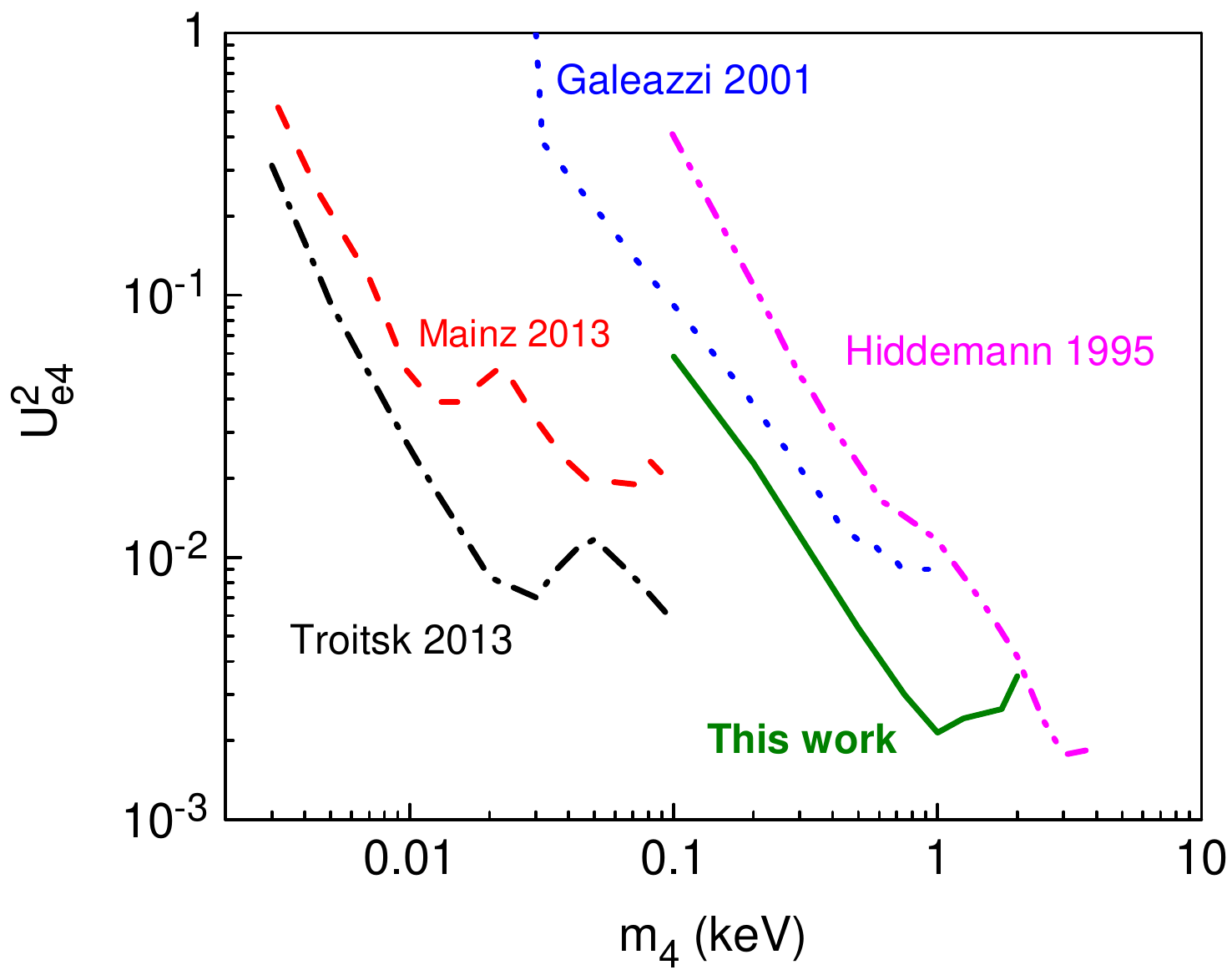}
	\end{center}
	\caption{Upper limits  on $U^2_{e4}$ at 95\% confidence level. Our result is shown by the solid curve. Published data are from~\cite{Kraus:2012he, Belesev:2012hx, Belesev2, Galeazzi:2001py, Hiddemann:1995ce}.  }
	\label{fig:limits}
\end{figure}

\paragraph{Acknowledgements.}
We are grateful to D. Gorbunov and  A. Lokhov for useful discussions.
This work was supported by the RFBR grant numbers 14-22-03069-ofi-m and  14-02-00570-a. IT also acknowledges the support of the Grant of President of Russian Federation for the leading scientific Schools of Russian Federation, NSh-9022-2016.2.

\end{document}